\documentclass[fleqn,twoside,amssymb]{article}
\usepackage{espcrc2}
\usepackage{graphicx}
\usepackage[T1]{fontenc}
\usepackage[latin1]{inputenc}
\usepackage[figuresright]{rotating}
\usepackage{amssymb}
\usepackage{amsfonts}
\usepackage{amsmath}

\title{Stochastic Background of Relic Scalar Gravitational Waves tuned by Extended Gravity}

\author{Mariafelicia De Laurentis\address{Dipartimento di Scienze Fisiche, Universit\'a
di Napoli {}`` Federico II'',\\ INFN Sez. di Napoli, Compl. Univ. di
Monte S. Angelo, Edificio G, Via Cinthia, I-80126, Napoli, Italy}, Salvatore Capozziello\addressmark}      
       
\begin{document}

\begin{abstract}
A stochastic background of relic gravitational waves is achieved
by  the so called adiabatically-amplified zero-point fluctuations
process derived from  early inflation.  It provides
a distinctive spectrum of relic gravitational waves. In the
framework of scalar-tensor gravity, we discuss the  scalar modes
of gravitational waves and  the primordial production of this
scalar component which is generated beside  tensorial one. Then we analyze different viable $f(R)$-gravities towards the
Solar System tests and stochastic gravitational waves background. The aim is to achive experimental bounds for the theory at local and
cosmological scales in order to select models capable of addressing the accelerating cosmological
expansion without cosmological constant but evading the weak field constraints. It is
demonstrated that viable $f(R)$-gravities under
consideration not only satisfy the local tests, but additionally,
pass the PPN-and stochastic gravitational waves bounds for
large classes of parameters.

\vspace{1pc}
\end{abstract}

\maketitle
\section{Introduction}
\label{uno}
The idea that General Relativity (GR) should be extended or corrected at
large scales (infrared limit) or at high energies (ultraviolet
limit) is suggested by several theoretical and observational
issues \cite{Mauro}. Quantum field theory in curved spacetimes, as well as the
low-energy limit of String/M theory, both imply semi-classical
effective actions containing  higher-order curvature invariants or
scalar-tensor terms. In addition, GR has been definitely tested
only at Solar System scales while it may show several shortcomings
if checked at higher energies or larger scales. Besides,  the
Solar System experiments are, up to now, not conclusive to
state that the only viable theory of gravity  is GR: for example,
the limits on PPN parameters should be greatly improved to fully
remove degeneracies.
Of course, modifying the gravitational action asks for several
fundamental challenges. These models can exhibit instabilities
 or ghost\,-\,like behavior, while, on the other hand, they have to be
matched with observations and experiments in the appropriate low
energy limit.
there exist cosmological solutions that give the
accelerated expansion of the universe at late times.  In addition, it has been
discovered that some stability conditions can lead to avoid ghost
and tachyon solutions. Furthermore there exist viable $f(R)$
models which satisfy both background cosmological constraints and
stability conditions and
results have been achieved in order  to place constraints on
$f(R)$ cosmological models by CMBR anisotropies and galaxy power
spectrum. Moreover, some of such
viable models lead to the unification of early-time inflation with
late-time acceleration. On the other hand, by considering $f(R)$-gravity in the low energy
limit, it is possible to obtain corrected gravitational potentials
capable of explaining  the flat rotation curves of spiral galaxies
or the  dynamics of galaxy clusters without considering huge
amounts of dark matter.
Furthermore, several authors have dealt with the weak field limit
of fourth order gravity, in particular considering the PPN limit
and the spherically symmetric solutions.
This great deal of work needs an essential issue to be pursued: we
need to compare experiments and probes at local scales (e.g. Solar
System) with experiments and probes at large scales (Galaxy,
extragalactic scales, cosmology) in order to achieve
self-consistent $f(R)$ models.
In order to constrain further viable $f(R)$-models, one could take
into account also  the stochastic background of gravitational
waves (GW) which, together with cosmic microwave background
radiation (CMBR), would  carry a huge amount of information on the
early stages of the Universe evolution.  In fact, if detected,
such a background could constitute a further probe for these
theories at very high red-shift \cite{tuning}. On the other hand,
a key role for the production and the detection of the relic
gravitational radiation background is played by the adopted theory
of gravity . This means that the effective
theory of gravity should be probed at zero, intermediate and high
redshifts to be consistent at all scales and not simply
extrapolated up to the last scattering surface, as in the case of
GR.

The aim of this report is to discuss the  PPN Solar-System
constraints and the GW stochastic background considering some
recently proposed $f(R)$ gravity models
\cite{Star,Hu} which satisfy both
cosmological and stability conditions mentioned above. Using the definition of
PPN-parameters $\gamma$ and $\beta$ in terms of $f(R)$-models
 and the definition of scalar GWs \cite{CCD}, we
compare and discuss if it is possible to search for parameter
ranges of  $f(R)$-models  working at Solar System and GW
stochastic background scale \cite{CDNO}.

\section{Extended Gravity}
\label{due}
A  action for scalar-tensor
gravity in a Brans-Dicke-like form which can be adopted also for $f(R)$-gravity once a suitable scalar field is defined,  is \cite{BD}
\begin{equation}
S=\int d^{4}x\sqrt{-g}\left[\varphi
R-\frac{\omega(\varphi)}{\varphi}g^{\mu\nu}\varphi_{;\mu}\varphi_{;\nu}-W(\varphi)+
\mathcal{L}_{m}\right]\label{eq: scalar-tensor2}
\end{equation}
By varying the action (\ref{eq: scalar-tensor2}) with respect to
$g_{\mu\nu}$, we obtain  the field equations
\begin{equation}
\begin{array}{c}
G_{\mu\nu}=-\frac{4\pi\tilde{G}}{\varphi}T_{\mu\nu}^{(m)}+\frac{\omega(\varphi)}{\varphi^{2}}(\varphi_{;\mu}\varphi_{;\nu}-\frac{1}{2}g_{\mu\nu}g^{\alpha\beta}\varphi_{;\alpha}\varphi_{;\beta})\\
\\+\frac{1}{\varphi}(\varphi_{;\mu\nu}-g_{\mu\nu}\square \varphi)+\frac{1}{2\varphi}g_{\mu\nu}W(\varphi)\end{array}\label{eq: einstein-general}\end{equation}
while the variation with respect to $\varphi$ gives the  Klein -
Gordon equation
\begin{eqnarray}
&&\square
\varphi=\frac{1}{2\omega(\varphi)+3}[-4\pi\tilde{G}T^{(m)}\nonumber\\ &&+2W(\varphi)+\varphi
W'(\varphi)+\frac{d\omega(\varphi)}{d\varphi}g^{\mu\nu}\varphi_{;\mu}\varphi_{;\nu}].\nonumber\\ \label{eq:
KG}\end{eqnarray} 
We are assuming physical units $G=1$, $c=1$ and
$\hbar=1$. $T_{\mu\nu}^{(m)}$ is the matter stress-energy tensor
and $\tilde{G}$ is a dimensional, strictly positive, gravitational
coupling  constant \cite{CCD,Capozz}. The Newton constant is
replaced by the effective coupling
\begin{equation}
G_{eff}=-\frac{1}{2\varphi},\label{eq: newton eff}\end{equation}
which is, in general, different from $G$. GR is
recovered for
\begin{equation}
\varphi=\varphi_0=-\frac{1}{2}.\label{eq: varphi}\end{equation}
\section{Gravitational waves from Extended Gravity}
\label{tre}
 In
order to study gravitational waves, we  assume first-order, small
perturbations in vacuum ($T_{\mu\nu}^{(m)}=0$). This means
\begin{equation}
g_{\mu\nu}=\eta_{\mu\nu}+h_{\mu\nu}\,,\qquad
\varphi=\varphi_{0}+\delta\varphi\label{eq:
linearizza}\end{equation}
and
\begin{equation}
W\simeq\frac{1}{2}\alpha\delta\varphi^{2}\Rightarrow
W'\simeq\alpha\delta\varphi\label{eq: minimo}\end{equation} for
the self-interacting, scalar-field potential. These assumptions
allow to derive the "linearized" curvature invariants
$\widetilde{R}_{\mu\nu\rho\sigma}$ , $\widetilde{R}_{\mu\nu}$ and
$\widetilde{R}$  which correspond to $R_{\mu\nu\rho\sigma}$ ,
$R_{\mu\nu}$ and $R$, and then the linearized field equations
\cite{CCD,Misner}
\begin{equation}
\begin{array}{c}
\widetilde{R}_{\mu\nu}-\frac{\widetilde{R}}{2}\eta_{\mu\nu}=-\partial_{\mu}\partial_{\nu}\Phi+\eta_{\mu\nu}\square \Phi\\
\\{}\square \Phi=m^{2}\Phi,\end{array}\label{eq: linearizzate1}\end{equation}
 where
\begin{equation}
\Phi\equiv-\frac{\delta\varphi}{\varphi_{0}}\,,\qquad
m^{2}\equiv\frac{\alpha\varphi_{0}}{2\omega+3}\,.\label{eq:
definizione}\end{equation}
 In particular, the
transverse-traceless (TT) gauge (see \cite{Misner}) can be
generalized to scalar-tensor gravity obtaining the total
perturbation of a GW incoming in the $z+$ direction in this gauge
as
\begin{eqnarray}
&&h_{\mu\nu}(t-z)=A^{+}(t-z)e_{\mu\nu}^{(+)}+A^{\times}(t-z)e_{\mu\nu}^{(\times)}\nonumber\\ &&+\Phi(t-z)e_{\mu\nu}^{(s)}.\label{eq:perturbazione totale}\end{eqnarray} The term
$A^{+}(t-z)e_{\mu\nu}^{(+)}+A^{\times}(t-z)e_{\mu\nu}^{(\times)}$
describes the two standard (i.e. tensorial) polarizations of a
gravitational wave arising from GR in the TT gauge
\cite{Misner}, while the term $\Phi(t-z)e_{\mu\nu}^{(s)}$ is the
extension of the TT gauge to the scalar case. This means that, in
scalar-tensor  gravity, the scalar field generates a third
component for the tensor polarization of GWs. This is because
three different  degrees of freedom are present of
\cite{CCD,SCF}, while only two are present in standard General
Relativity.
\section{Stochastic background of relic scalar GWs}
\label{quattro}

Then,  for a purely scalar gravitational wave,
the metric perturbation is \cite{CCD,SCF,tuning}
 \begin{equation}
h_{\mu\nu}=\Phi \, e_{\mu\nu}^{(s)} \;. \label{eq: perturbazionescalare}
\end{equation}
 The stochastic background of scalar gravitational waves can be
described in terms of the scalar field $\Phi$ and characterized by
a dimensionless spectrum (see the analogous definition for
tensor modes in
\cite{CCD})
\begin{equation}
\Omega_{sgw}(f)=\frac{1}{\rho_{c}}\frac{d\rho_{sgw}}{d\ln
f} \;,\label{eq:spettro}
\end{equation} where
 \begin{equation}
\rho_{c}\equiv\frac{3H_{0}^{2}}{8\pi G}\label{eq: densita'
critica}\end{equation}
is the (present) critical energy density of
the universe, $H_0$ is the  Hubble parameter today, and
$d\rho_{sgw}$ is the energy density of the scalar
gravitational radiation in the frequency interval $ \left( f,
f+df \right)$. We are now using standard units.
Now it is possible to write an
expression for the energy density of the stochastic scalar relic
gravitons background in the angular frequency interval
$(\omega,\omega+d\omega)$ as
\begin{eqnarray}
&&d\rho_{sgw}=2\hbar\omega\left(\frac{\omega^{2}d\omega}{2\pi^{2}c^{3}}\right)N_{\omega}=\nonumber\\ &&
\frac{\hbar
H_{dS}^{2}H_{0}^{2}}{4\pi^{2}c^{3}}\frac{d\omega}{\omega}=\frac{\hbar
H_{dS}^{2}H_{0}^{2}}{4\pi^{2}c^{3}} \frac{df}{f}\,,\label{eq: de
energia}\end{eqnarray}
where $f$, as above, is the frequency in
standard comoving time. Eq.~(\ref{eq: de energia}) can be
rewritten in terms of the critical and de Sitter  energy
densities
\begin{equation}
H_{0}^2 =\frac{8\pi G\rho_{c}}{3c^{2}}\,,\qquad
H_{dS}=\frac{8\pi G\rho_{dS}}{3c^{2}} \;.
\end{equation}
Introducing the Planck density ${\displaystyle
\rho_{Planck}=\frac{c^{5}}{\hbar G^{2}}}$, the spectrum is given
by
\begin{equation}
\Omega_{sgw}(f)=\frac{1}{\rho_{c}}\frac{d\rho_{sgw}}{d\ln
f}=\frac{f}{\rho_{c}}\frac{d\rho_{sgw}}{df}=\frac{16}{9}
\frac{\rho_{dS}}{\rho_{Planck}} \;.\label{eq:spettrogravitoni}
\end{equation}
At this point,  some  comments
are in order. First, the calculation works for a
simplified model that does not include the matter-dominated era.
If the latter is included, the redshift at the equivalence
epoch has to be considered. Taking into account Ref.~\cite{CCD,SCF} one
gets
\begin{equation}
\Omega_{sgw}(f)=\frac{16}{9}\frac{\rho_{dS}}{\rho_{Planck}}(
1+z_{eq})^{-1}
\label{eq:spettrogravitoniredshiftato}
\end{equation}
for the waves which,
at the epoch in which the universe becomes matter-dominated, have
a frequency higher than $H_{eq}$, the Hubble parameter at
equivalence. This situation corresponds to frequencies
$f>(1+z_{eq})^{1/2}H_{0}$ today. The redshift correction in
eq.~(\ref{eq:spettrogravitoniredshiftato}) is needed since the present value of
the
Hubble parameter $H_{0}$ would be different  without a matter-dominated
contribution. At lower frequencies, the spectrum is
given by 
\begin{equation}
\Omega_{sgw}(f)\propto f^{-2}.\label{eq:spettrobassefrequenze}
\end{equation} 
Nevertheless, since the spectrum falls off as $ f^{-2}$ at low
frequencies, today at {\em LIGO/VIRGO} and {\em LISA}
frequencies, one gets
\begin{equation} \Omega_{sgw}(f)h_{100}^{2}<2.3\times
10^{-12} \;,\label{eq:limitespettroWMAP}
\end{equation}
where $h_{100}=H_0/\left( 100 \; \mbox{km}\cdot \mbox{s}^{-1}
\cdot \mbox{Mpc}^{-1} \right)$.  It is interesting to calculate
the  corresponding strain at $ f \sim 100$Hz, where
interferometers such as {\em VIRGO} and {\em LIGO} achieve maximum
sensitivity. The well known equation for the characteristic
amplitude adapted to the scalar component
of gravitational waves
\begin{equation}
\Phi_{c}(f)\simeq1.26\times
10^{-18}\left(
\frac{1 \, \mbox{Hz}}{f} \right)\sqrt{h_{100}^{2}\Omega_{sgw}(f)}
\;, \label{eq:legameampiezza-spettro}
\end{equation}
can be used to obtain
\begin{equation}
\Phi_{c}\left( 100\, \mbox{Hz} \right) < 2\cdot 10^{-26}\;.
\label{eq:limiteperlostrain}
\end{equation}
Then, since we expect a sensitivity of the order of $10^{-22}$ for
the above interferometers at $f \sim 100$~Hz, we need to gain four
orders of magnitude. Let us analyze the situation also at lower
frequencies. The sensitivity of the {\em VIRGO} interferometer is of the
order of $10^{-21}$ at $f\sim 10$~Hz and in that case it is
\begin{equation}
\Phi_{c} \left( 10 \, \mbox{Hz} \right) < 2\cdot 10^{-25} \;.
\label{eq:limiteperlostrain2}\end{equation}
The sensitivity of the
{\em LISA} interferometer
will be of the order of $10^{-22}$ at $ f \sim 10^{-3} $~Hz and in
this case it is
\begin{equation}
\Phi_{c} \left( 10^{-3} \mbox{Hz} \right) <2\cdot 10^{-21} \;.
\label{eq:limiteperlostrain3}
\end{equation}
This means that a stochastic background of
relic scalar gravitational waves could, in  principle, be detected by
the {\em LISA} interferometer \cite{BCDF}.

\section{$f(R)$-gravity constrained by PPN parameters and stochastic background of GWs}
\label{cinque}
\begin{figure}[]
\includegraphics[scale=0.3]{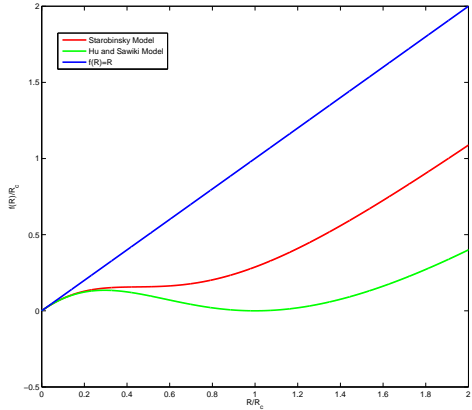}
\caption{Plots of  $f(R)$ models as function of
$\frac{R}{R_{c}}$.  Model $f_{HS}$ in Eq. (\ref{eq:HS}) with $n=1$ and
$\lambda=2$ (green line). Model $f_{S}$ in Eq.(\ref{eq:STAROBINSKY})
with $p=2$, $\lambda=0.95$ (red line). 
We also plot  $f(R)=R$ (blu line) to see whether or not
the stability condition $f_{,R}>0$ is violated. } \label{fig:f(R)}
\end{figure}

A Brans-Dike-like theory with $\omega=0$ is dinamically equivalent to an $f(R)$-gravity, so
the bounds coming from the interferometric ground-based (VIRGO, LIGO) and space (LISA)
 experiments could constitute a further probe for  gravity if matched with bounds at other scales to achieve experimental bounds for the theory at local and cosmological scales
. For our aims consider a class of $f(R)$ models which do not contain
cosmological constant and  are explicitly designed to satisfy
cosmological and Solar-System constraints in given limits of the
parameter  space. In practice, we choose a class of functional
forms of $f(R)$ capable of matching, in principle, observational
data. Firstly, the
cosmological model  should reproduce the  CMBR constraints in the
high-redshift regime (which agree with the presence of an
effective cosmological constant). Secondly, it should give rise to
an accelerated expansion, at low redshift, according to the
$\Lambda$CDM model. Thirdly, there should be sufficient degrees of
freedom in the parameterization to encompass low redshift
phenomena (e.g. the large scale structure) according to the
observations. Finally, small deviations from GR
should be consistent with Solar System tests.
All these
requirements suggest that we can assume the  limits
\begin{equation}
\lim_{R\rightarrow\infty}f(R)={\rm constant},
\end{equation}
\begin{equation}
\lim_{R\rightarrow0}f(R)=0,
\end{equation}
which are satisfied by a general class of broken power law models,
proposed in \cite{Hu}, which are
\begin{equation}
f_{HS}(R)=R-\lambda
R_{c}\frac{\left(\frac{R}{R_{c}}\right)^{2n}}{\left(\frac{R}{R_{c}}\right)^{2n}+1}
\label{eq:HS}\end{equation}
where $m$ is a  mass scale and
$c_{1,2}$ are dimensionless parameters.
Besides, another viable class of models  was proposed in
\cite{Star}

\begin{equation}
f_{S}(R)=R+\lambda
R_{c}\left[\left(1+\frac{R^{2}}{R_{c}^{2}}\right)^{-p}-1\right]\,.\label{eq:STAROBINSKY}\end{equation}
Since $f(R=0)=0$, the cosmological constant has to disappear in a
flat spacetime. The parameters  $\{n$, $p$, $\lambda$, $R_{c}\}$
are constants which should be determined by experimental bounds.
In
Fig.(\ref{fig:f(R)}), we have plotted the selected models
as function of ${\displaystyle \frac{R}{R_{c}}}$  for suitable
values of $\{p,n,\lambda\}$ .

The above models can be constrained at Solar System level by
considering the PPN formalism. This approach is extremely
important in order to test gravitational theories and to compare
them with GR.  One can
derive the PPN-parameters $\gamma$ and $\beta$ in terms of a
generic analytic function $f(R)$  and its derivative

\begin{equation}
\gamma-1=-\frac{f''(R)^{2}}{f'(R)+2f''(R)^{2}}\,,\label{eq:PPNgamma}\end{equation}

\begin{equation}
\beta-1=\frac{1}{4}\left[\frac{f'(R)\cdot
f''(R)}{2f'(R)+3f''(R)^{2}}\right]\frac{d\gamma}{dR}\,.\label{eq:PPNbeta}\end{equation}
These quantities have to fulfill the constraints  coming from the
Solar System experimental tests. They are
the perihelion shift of Mercury , the Lunar Laser
Ranging , the upper limits coming from the Very
Long Baseline Interferometry (VLBI) and the
results obtained from the Cassini spacecraft mission in the delay
of the radio waves transmission near the Solar conjunction.

\begin{itemize}
\item Mercury perihelion Shift $\left|2\gamma-\beta-1\right|<3\times10^{-3}$
\item  Lunar Laser Ranging $4\beta-\gamma-3=(0.7\pm1)\times10^{-3}$
\item Very Long Baseline Interferometer $\left|\gamma-1\right|<4\times10^{-4}$
\item Cassini Spacecraft $\gamma-1=(2.1\pm2.3)\times10^{-5}$
\end{itemize}
Take into account the above  $f(R)$-models,specifically, we have investigated
the values or the ranges of parameters in which they match the above
Solar-System experimental constraints. In other
words, we use these models to search  under what circumstances it
is possible to significantly address cosmological observations by
$f(R)$-gravity and, simultaneously, evade the local tests of
gravity \cite{CDNO}.
At this point, using the above LIGO, VIRGO and LISA upper bounds,
calculated for the characteristic amplitude of GW scalar
component, let us test the  $f(R)$-gravity models. 
Taking into account the
discussion in Sec. \ref{due},  we have that the GW scalar component is
derived considering
 \begin{eqnarray}
&&\Phi=-\frac{\delta\sigma}{\sigma_{0}}\,, \label{eq:PHI} \qquad
\sigma=-\ln(1+f'(R))=\ln
f'(A)\,,\nonumber\\ &&\qquad\delta\sigma=\frac{f''(A)}{1+f'(A)}\delta A\,.
\end{eqnarray}
 Finally we obtain a good  sets of parameters for the  Starobinsky model PPN vs GW-stochastic: 
\begin{itemize}
\item $\frac{R}{R_c}=3.38$\; , $p=1$\; , $\lambda=2$
\item $\frac{R}{R_c}=\sqrt{3}$\; , $p=2$\; , $0.944<\lambda<0.966$
\end{itemize}
and for Hu and Sawiki model
\begin{itemize}
\item $\frac{R}{R_c}=\sqrt{3}$\; , $n=2$\; , $\lambda>\frac{8}{3\sqrt{3}}$
\end{itemize}
such sets of parameters are not in conflict with
bounds coming from the cosmological stochastic background of GWs and, 
some sets  reproduce quite well both the PPN upper limits and the constraints on the scalar 
component amplitude of GWs. 
The results indicate that self-consistent models could be achieved comparing
experimental data at very different scales without extrapolating results obtained only at a given scale \cite{CDNO}.
\section{Conclusions}
We have investigated the possibility that some
viable $f(R)$ models could be constrained considering both Solar
System experiments and upper bounds on the stochastic background
of gravitational radiation. Such bounds   come from
interferometric ground-based (VIRGO and LIGO) and space (LISA)
experiments. The underlying philosophy is to show that the $f(R)$
approach, in order to describe consistently  the observed
universe, should  be tested at very different scales, that is at
very different redshifts. In other words, such a proposal could
partially contribute to remove the unpleasant degeneracy affecting
the wide class of dark energy models, today on the ground.

Beside the request to evade the Solar System tests, new methods
have been recently proposed to investigate the evolution and the
power spectrum of cosmological perturbations in $f(R)$ models.
 The investigation of stochastic background, in
particular of the scalar component of GWs coming from the $f(R)$
additional degrees of freedom, could acquire, if revealed by the
running and forthcoming experiments,  a fundamental importance to
discriminate among the various  gravity theories \cite{tuning}.
These data (today only upper bounds coming from simulations) if
combined with Solar System tests, CMBR anisotropies, LSS, etc.
could greatly help to achieve a self-consistent cosmology
bypassing the shortcomings of $\Lambda$CDM model.

Specifically, we have taken into account two broken power law
$f(R)$ models fulfilling the main cosmological requirements which
are to match the today observed accelerated expansion and the
correct behavior in early epochs. 
 We have taken into account the results of the
main  Solar System  current experiments. Such results give upper
limits on the PPN parameters which any self-consistent theory of
gravity should satisfy at local scales. Starting from these, we
have selected the $f(R)$ parameters fulfilling the tests. As a
general remark, all the functional forms chosen for $f(R)$ present
sets of parameters capable of matching the two main PPN
quantities, that is $\gamma_{exp}$ and $\beta_{exp}$. This means
that, in principle, extensions of GR are not {\it a priori}
excluded as reasonable candidates for gravity theories. To
construct such extensions, the reconstruction method developed in
 may be applied.

The interesting feature, and the main result of this paper, is
that such sets of parameters are not in conflict with bounds
coming from the cosmological stochastic background of GWs. In
particular, some sets of parameters  reproduce quite well both the
PPN upper limits and the constraints on the scalar component
amplitude of GWs \cite{CDNO,BCDF}.

Far to be definitive, these preliminary results indicate that
self-consistent models could be achieved comparing experimental
data at very different scales without extrapolating results
obtained only at a given scale.

\end{document}